\documentclass[manuscript]{aastex}
\usepackage{aas_macros,url,natbib,amssymb,amsmath,CJK}

\shorttitle{A search of reactivated comets}
\shortauthors{Ye}

\begin{document}
\begin{CJK*}{UTF8}{gbsn}

\title{A search of reactivated comets}

\author{Quan-Zhi Ye (叶泉志)}
\affil{Department of Physics and Astronomy, The University of Western Ontario, London, Ontario N6A 3K7, Canada}
\affil{Astronomy Department, California Institute of Technology, Pasadena, CA 91125, U.S.A.}
\affil{Infrared Processing and Analysis Center, California Institute of Technology, Pasadena, CA 91125, U.S.A.}
\email{qye@caltech.edu}

\begin{abstract}
Dormant or near-dormant short-period comets can unexpectedly regain ability to eject dust. In many known cases, the resurrection is short-lived and lasts less than one orbit. However, it is possible that some resurrected comets can remain active in latter perihelion passages. We search the archival images of various facilities to look for these ``reactivated'' comets. We identify two candidates, 297P/Beshore and 332P/Ikeya-Murakami, both of which were found to be inactive or weakly active in the previous orbit before their discovery. We derive a reactivation rate of $\sim0.007~\mathrm{comet}^{-1}~\mathrm{orbit}^{-1}$, which implies that typical short-period comets only become temporary dormant for less than a few times. Smaller comets are prone to rotational instability and may undergo temporary dormancy more frequently. Next generation high-cadence surveys may find more reactivation events of these comets.
\end{abstract}

\section{Introduction}

It is well known that the brightness of comets can undergo large, seemingly random fluctuation. Fluctuation involving short-term increases in activity (comet \textit{outbursts}) are distinctly noticeable and have attracted regular interests. It is suggested that comets can even resurrected from dormant\footnote{For definition of dormant comets, see, e.g. \citet{Weissman2002}, for discussion.} or near-dormant state \citep{1990A&A...237..524R}. In many known cases, the resurrection is short-lived; but it has been speculated that the resurrection can be long-lived -- i.e. comets will show activities in their proceeding perihelion passages just as normal comets \citep{1987A&A...187..906K,1990Icar...86...82K}. However, the inventory of these \textit{reactivated~comets} is virtually uncharted largely due to the difficulty to identify their inactive or weakly active progenitors.

The ever-increasing effort from various near-Earth object (NEO) surveys since the late 1990s has provided an excellent source of data to explore temporally variable phenomena such as reactivated short-period comets. As a starting point, it is useful to find short-period comets that have been in dormant or near-dormant state recently. Here, we present a search of the reactivated comets by examining pre-discovery data of known comets.

\section{Methodology and Results}

Since most NEO surveys started in the late 1990s, we focus at comets detected no earlier than $2000+5=2005$ (where 5 is the typical orbital period for short-period comets in years) if we want to include at least one pre-discovery orbit of the comet. We also check for cometary activity in the proceeding orbit after the orbit of discovery. We specifically exclude active asteroids that are the list compiled by \citet{2015aste.book..221J}. As the time of the writing, there are 40 comets satisfying these criteria (Table~\ref{tbl:all}).

We then searched the archival images provided by the SkyMorph service \citep[\url{http://skyview.gsfc.nasa.gov/skymorph/skymorph.html};][]{Lawrence1998}, the Solar System Object Search service hosted at Canadian Astronomy Data Centre \citep[CADC; \url{http://www.cadc-ccda.hia-iha.nrc-cnrc.gc.ca/en/ssois/index.html};][]{Gwyn2012b}, as well as the Catalina Sky Survey \citep[CSS;][]{2016DPS....4840501C}, for pre-discovery images of each comet in Table~\ref{tbl:all}. Images are selected based on the predicted position and brightness of the comets \citep[$m_\mathrm{T}<21$ for NEO survey images following][, and $m_\mathrm{T}<25$ for other images, where $m_\mathrm{T}$ is the total magnitude of the comet]{Jedicke2015}. The only comet excluded from this procedure is 332P/Ikeya-Murakami which extensive archival data search had been conducted by \citet{Hui2016a}. 

We found and retrieved pre-discovery images for a total of six comets. Images are reduced using the fourth U.S. Naval Observatory CCD
Astrograph Catalog \citep[UCAC4;][]{2013AJ....145...44Z}. We identified previously unreported pre-discovery observation for 3 comets (297P/Beshore, 317P/WISE and 336P/McNaught); while for other images, we estimated the limiting magnitude of the images to the nearest 0.5 mag using the faintest visible stars. Details of the pre-discovery (non-)detections are summarized in Table~\ref{tbl:can}. The involved facilities are summarized in Table~\ref{tbl:fac}.

Individual comets are discussed below:

\paragraph{213P/Van Ness} (semimajor axis $a=3.43$~AU, eccentricity $e=0.38$, inclination $i=10.2^\circ$) was discovered in September 2005 during an outburst of the comet \citep{vanNess2005b}. It was found to have split during the subsequent return in 2011, but the actual split might take place just a few days before the discovery in 2005 \citep{Hanayama2011a}. We found 7 sets of pre-discovery images taken by the Near-Earth Asteroid Tracking (NEAT) survey in 2002--2003, where the comet was $\sim 4$~AU from the Sun. The comet was predicted to be $V\sim 20$ around these times, though it is likely an exaggerated value since most observations are made after the outburst/fragmentation. The JPL orbit solution is not suitable for us due to the complication arises from the comet's history of fragmentation, hence, we calculated the orbit and covariance matrix of the comet using the FindOrb package (\url{http://www.projectpluto.com/find_orb.htm}) based on the pre-fragmentation observations taken in August 2005, available from the Minor Planet Center database (\url{http://www.minorplanetcenter.net/db_search}). The pre-outburst orbit is presented alongside with the most recent JPL orbit in Table~\ref{tbl:213p-orb}.

According to the calculation, the NEAT images are wide enough to cover the entire uncertainty ellipse. The images were blinked to reveal moving objects. We searched the entire images and specifically look for objects that match the motion of 213P/Van Ness which none is found. Since the depth of the images only barely reach the predicted $m_\mathrm{T}$ which is likely already inflated as aforementioned, we are only able to conclude that the comet was unlikely to as bright as expected in April--May 2003.

\paragraph{266P/Christensen} ($a=3.53$~AU, $e=0.34$, $i=3.4^\circ$) was discovered in October 2006 \citep{Christensen2006bo}. We found 2 sets of pre-discovery images taken in 2001, which neither is deep enough to reach the predicted $m_\mathrm{T}$. Nevertheless, we blinked the images to search for the comet but nothing was found.

\paragraph{297P/Beshore} ($a=3.48$~AU, $e=0.31$, $i=10.3^\circ$) was discovered in May 2008 at a heliocentric distance of $r_\mathrm{H}=2.43$~AU at an unusually bright 14th magnitude \citep{Beshore2008x}. At typical cometary brightening rate ($\propto r_\mathrm{H}^{-4}$), 297P/Beshore would have been brighter than most NEO survey limits ($V\sim19$) since early 2007. The position of the comet was scanned no less than 5 times within three months before discovery, in which the comet would have been 15--16 mag. This strongly suggests that 297P/Beshore was discovered following a large outburst.

\begin{figure}
\includegraphics[width=\textwidth]{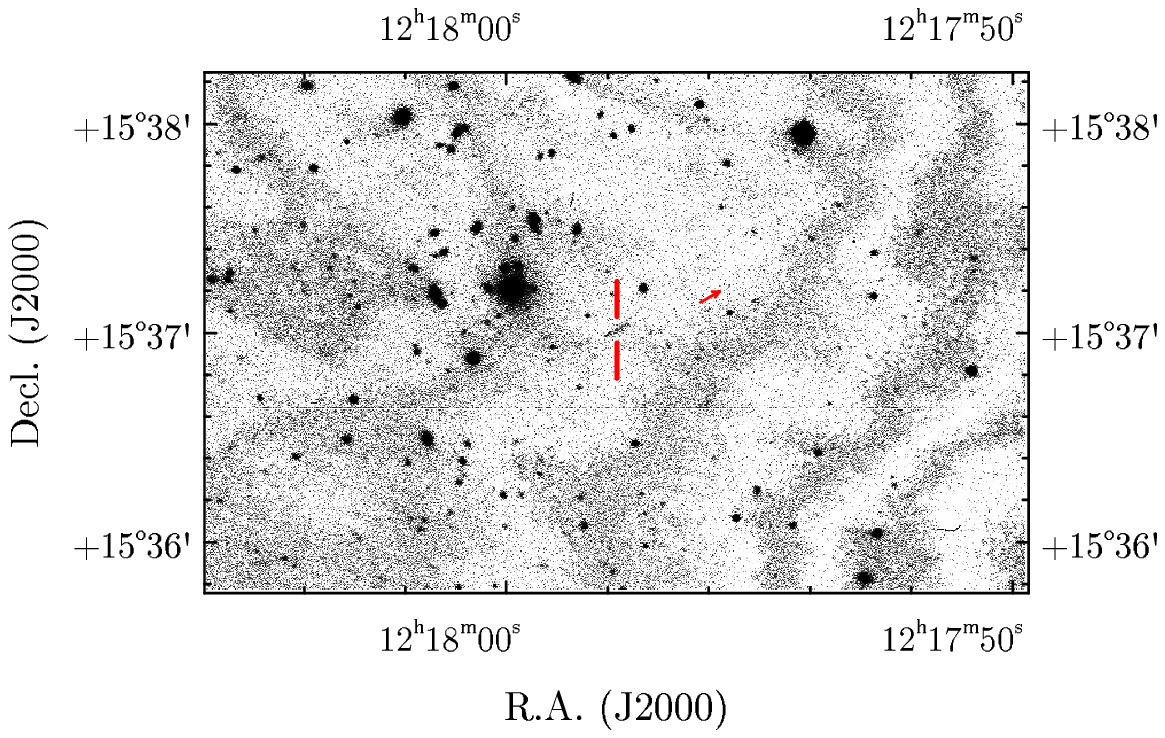}
\caption{Pre-discovery image of 297P/Beshore (center), taken by the Isaac Newton Telescope on 2001 March 20. The comet (highlighted streak at the center of the image) was trailed due to the motion of the comet and the long exposure. The red arrow marks the predicted motion of the comet computed from JPL orbit \#33.}
\label{fig:297p-int}
\end{figure}

We found 12 sets of pre-discovery images in 2001--2008 of which nothing is found in all but one of them. The comet is readily visible (as a short streak) in the 900~s exposure taken by the Wide Field Camera on the Isaac Newton Telescope (INT) on 2001 Mar. 20, of which the comet was at $r_\mathrm{H}=2.65$ (Figure~\ref{fig:297p-int}). The image retrieved from the INT archive has moderate gradient and is first corrected by fitting and subtracting the background with a high order polynomial function before photometric reduction. The trailing loss is then corrected for photometric measurement of the comet. For other sets of images, we blinked them to look for moving objects that match the motion of the comet and found nothing. The updated orbit, alongside with the pre-discovery INT observations included, is presented in Table~\ref{tbl:297p-orb}.

297P/Beshore was measured to be $V\sim 22$ in the INT data. This can be used to constrain the nucleus size, by

\begin{equation}
\label{eq:d_n}
D_\mathrm{N} = (3\times10^8)~p^{-\frac{1}{2}}~10^{0.2(m_\odot-M_\mathrm{N})}
\end{equation}

\noindent whereas $M_\mathrm{N} = m_\mathrm{N} - 5\log{(r_\mathrm{H} \varDelta)} - \alpha \beta$ is the absolute cometary nuclear magnitude, of which $m_\mathrm{N}>22$ is the apparent nuclear magnitude, $r_\mathrm{H}$, $\varDelta$ is the heliocentric distance and the geocentric distance of the comet in AU, respectively, $\alpha$ is the phase angle of the comet in degree, $\beta=0.04~\mathrm{mag/deg}$ and $p=0.04$ is the phase coefficient and albedo, respectively \citep{2004come.book..223L}, and $m_\odot=-26.8$ is the apparent magnitude of the Sun. We can hereby derive $D_\mathrm{N}\lesssim1$~km. This indicates that the comet was either inactive or very weakly active at the time of the observation, or has a sub-km-sized nucleus. We will revisit this issue in the discussion. The non-detection in the CSS and Siding Spring Survey (SSS) images in May 2007 and March 2008 provided further support to the conclusion that the comet was discovered following a large outburst in 2008.

The comet was recovered in early 2014 at 20th magnitude without any information about its appearance \citep{2014CBET.3813....1D}. However, the comet was about 2 mag brighter than the brightness extrapolated from the 2001 data, indicating that the comet was more active than in 2001.

\paragraph{302P/Lemmon-PANSTARRS} ($a=4.27$~AU, $e=0.23$, $i=6.0^\circ$) was discovered in July 2007 \citep{Bolin2014}. It has only 1 set of pre-discovery images found and is not deep enough to allow any conclusion.

\paragraph{317P/WISE} ($a=2.93$~AU, $e=0.59$, $i=10.8^\circ$) was discovered in May 2010 \citep{Scotti2010bf}. We found 5 sets of pre-discovery images found during its last undetected perihelion passage in 2005. The comet was visible on the SSS images taken on 2005 July 28, being about the same brightness as prediction. It is likely a low activity comet \citep{2016Icar..264...48Y} rather than a reactivated comet.

\paragraph{332P/Ikeya-Murakami} ($a=3.09$~AU, $e=0.49$, $i=9.4^\circ$) was discovered in November 2010 at $r_\mathrm{H}=1.60$~AU following an apparent outburst \citep{2014ApJ...787...55I}. \citet{Hui2016a} searched a set of archival data in 2003--2005 and placed an upper limit of nucleus diameter $D_\mathrm{N}<1$~km, which led them concluded that the comet was largely inactive prior to its 2010 perihelion passage. Independent observation with the Hubble Space Telescope has placed a tighter limit of $D_\mathrm{N}<0.55$~km of the pre-outburst progenitor \citep{2016ApJ...829L...8J}.

The comet was recovered in late 2015 with realization that it had split into a few dozens of fragments \citep[e.g.][]{2016ApJ...827L..26K}. Observations suggested that sublimation-driven mass loss are still ongoing on these fragments and some of the fragments continue to split.

\paragraph{336P/McNaught} ($a=4.81$~AU, $e=0.45$, $i=18.6^\circ$) was discovered in April 2006 \citep{2006IAUC.8699....1M} and has 5 sets of pre-discovery images found. The comet was visible on the NEAT images taken on 1996 Aug. 11, being about 1 mag fainter than prediction but still 4 mag brighter than bare nucleus brightness. We therefore concluded that the comet was active in its 1996 perihelion.

\section{Discussion}

We identified comets 297P/Beshore and 332P/Ikeya-Murakami as plausible reactivated comets.

It is interesting that the reactivation of both comets are marked by large outbursts: 297P/Beshore had brightened by at least 5--6 mag; for 332P/Ikeya-Mukarami it is not known how much it had brightened, but the comet reportedly lost $4\%$ of its mass \citep{2016ApJ...829L...8J}, comparable to the well-studied mega-outburst exhibited by 17P/Holmes in 2007 \citep{2011ApJ...728...31L}. The repeated activity of 297P/Beshore and 332P/Ikeya-Murakami into their next perihelion passages suggests that the mass-loss mechanism is re-triggered when the comet approaches the Sun, and that the activity would not be quickly shut down by aging and environmental effects, consistent with sublimation-driven activity. If so, a large nucleus disturbance (or even a disruption of the nucleus) is likely required to break the mantle that seal off the volatile and is consistent with the large outbursts observed at the reactivation of both comets. For 332P/Ikeya-Murakami, \citet{2016ApJ...829L...8J} suggested that rotational excitation as a likely driving force, while for 297P/Beshore no studies have been published as of December 2016. Other mechanisms such as asteroid impact, tidal and thermal stress, as well as amorphous ice crystallization are also known to cause nucleus disturbance or disruption. However, the occurrence of asteroid impact for a typical km-wide short-period comet is $\sim 10^{-3}~\mathrm{comet^{-1}~orbit^{-1}}$ \citep{2002EM&P...88..211B}, which is an unlikely event; tidal and thermal stress requires the comet to be sufficiently close to a giant planet or the Sun. Crystallization has been proposed to be the outburst trigger for the cases of 17P/Holmes \citep{2011ApJ...728...31L} and 332P/Ikeya-Murakami \citep{2014ApJ...787...55I}, but it is unclear if amorphous ice does exist on cometary surfaces.

\citet{Hui2016a} suggested a potential linkage between 332P/Ikeya-Murakami and P/2010 B2 (WISE). Such linkage, if real, would imply previous fragmentation of the progenitor of the two comets and provide an evolutionary sketch of a cascading fragmentation of a comet. We tested this idea on 297P/Beshore and searched for objects in similar orbits. The closest comet is P/2005 JN (Spacewatch) with the \citet{1963SCoA....7..261S}'s $D$-criterion $D_\mathrm{SH}=0.16$, while the closest asteroid is 2014 JO, with $D_\mathrm{SH}=0.10$, but neither is as close as the 332P-B2 pair ($D_\mathrm{SH}=0.04$). This may be considered as another evidence, in addition to the fact that 332P/Ikeya-Murakami is observed to have fragmented while 297P/Beshore is not, that the evolutionary history of the two comets is different.

Is it possible that the two comets are just smaller (sub-km), moderately active comets which activity would not be noticeable without a large outburst? Though fully active ($100\%$ active surface) sub-km comets are quickly eliminated by rotational instability, moderately active ($\sim1\%$ active surface) sub-km comets are less prone to such effects and can survive up to $\sim10^2$~yr \citep{2016ApJ...829L...8J}, making them more likely to be detected though probably without being recognized as a comet. To answer this question, we consider the comet \textit{recognizability}, defined by the sign of $M_\mathrm{T}-M_\mathrm{N}$; and the rotational instability of comets, defined by \citet{1997EM&P...79...35J}. Here $M_\mathrm{T}$ is the absolute total magnitude of comets, derived using the relation determined by \citet{2008LPICo1405.8046J}, assuming the comet activity is driven by water ice sublimation; $M_\mathrm{N}$ is calculated using the aforementioned relation embedded in Eq.~\ref{eq:d_n} assuming a geometric albedo of 0.04, $r_\mathrm{H}=\varDelta=1$~AU, and $\alpha=0^\circ$. The main idea behind this definition is that comets will likely to be recognized when they produce enough dust that exceed the nuclear brightness. The local sublimation rate is derived from the sublimation energy balance equation \citep{1979M&P....21..155C}. For rotational instability equation, we followed the parameters discussed and adopted in \citet{2016ApJ...829L...8J} except taking the moment-arm $k_\mathrm{T}\sim0.01$ \citep{Belton2014}. Comets that can be detected need to have the disruption timescale, $\tau_\mathrm{s}$, longer than the characteristic timescale that observers can find them, $\tau_\mathrm{o}$, of which we take $\tau_\mathrm{o}\sim20$~yr. The physical meaning of $\tau_\mathrm{o}$ is that, if a comet is disrupted before it has completed enough orbits that it will be detected in any of these orbits, we would not know it had existed.

\begin{figure*}
	\includegraphics[width=\textwidth]{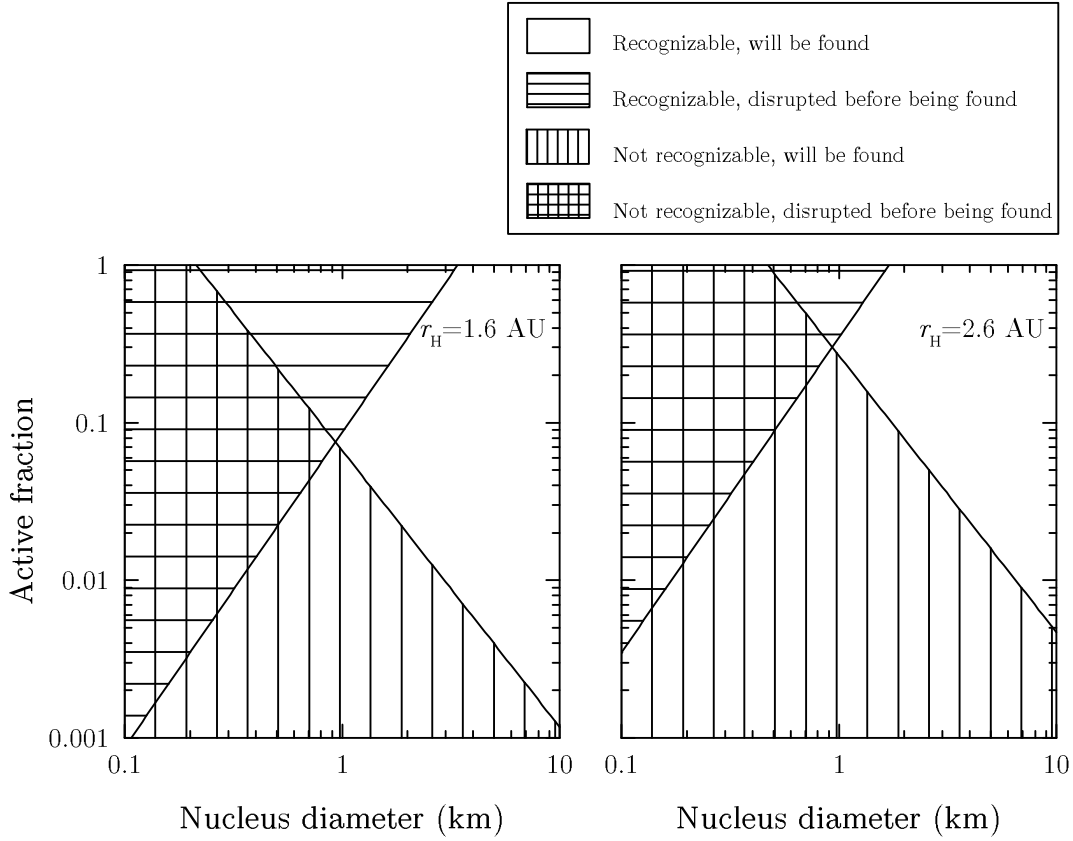}
	\caption{Recognizability and rotational stability of comets at different heliocentric distance, $r_\mathrm{h}=1.6$~AU (appropriated to the pre-discovery detection of 332P/Ikeya-Murakami reported by \citet{Hui2016a}), and $r_\mathrm{h}=2.6$~AU (appropriated to the pre-discovery detection of 297P/Beshore reported in this work), as a function of the fraction of active surface and nucleus size. A comet is considered detectable when $M_\mathrm{T}<M_\mathrm{N}$ and vice versa, of which $M_\mathrm{T}$ is derived assuming the comet activity is driven by water ice sublimation. The rotational stability is calculated using \citet{1997EM&P...79...35J} taking disruption timescale to be 20~yr.}
	\label{fig:small-comet}
\end{figure*}

As shown in Figure~\ref{fig:small-comet}, the two indicators -- recognizability and rotational instability -- divide the graph into four quadrants: (i) comets that can be recognized as such and will be found; (ii) comets that can be recognized as such but will be disrupted before being found; (iii) low activity comets that cannot be recognized but will be found as asteroids; and (iv) low activity comets that cannot be recognized as such, and will be disrupted before being found.

The figure draws several interesting conclusions:

\begin{itemize}
 \item Most 1 km-sized comets at a few AU are difficult to recognized, unless they are very active (active fraction $\gg10\%$) or are in outbursts.
 \item Rotational instability quickly depletes active sub-km-sized comets before current NEO survey can find them. The predicted observable size population peaks around 1~km, which are in line with the observation \citep{Snodgrass2011}.
\end{itemize}

This provides further support to our previous conclusion that 297P/Beshore and 332P/Ikeya-Murakami had been weakly active before reactivation, since both comets have resided in the inner solar system for at least a few $10^2$~yr and are not completely disrupted\footnote{Besides \citet{Hui2016a}'s work on 332P/Ikeya-Murakami, \citet{Tancredi2014} also maintain an analysis of the dynamical evolution of comets at \url{http://www.astronomia.edu.uy/Criterion/Comets/Dynamics/table_num.html}, retrieved 2016 Dec. 18.}.

The result leads to a more generic question: how often do comets reactivate? By querying the JPL database, we found that there are about 100 comets that (a) have been observed at two orbits from 2005 to now; and (b) had been observed at their last orbit before 2005 (i.e. these comets have been observed for 20~years and have completed 3 orbits). The rate of reactivation is therefore $2/100/3\sim0.007~\mathrm{comet^{-1}~orbit^{-1}}$. This rate is equivalent to the frequency of comets entering temporary dormancy, as comets must become dormant before reactivate. Since short-period comets are only physically active for a few hundred orbits \citep{2002Icar..159..358F}, this number seems to suggest that typical short-period comets likely only become temporary dormant for less than a few times before their ultimate end, assuming no individual differences. On the other hand, if rotational excitation turns out to be the dominant mechanism in reactivating comets, temporary dormant comets will be dominated by smaller comets, while larger comets do not or very rarely become temporary dormant.

\section{Summary}

We conducted a search to look for short-period comets that are reactivated from dormant or near-dormant stage and are able to sustain their activity into their latter orbits. Comets 297P/Beshore and 332P/Ikeya-Murakami are identified as such comets. Both comets were discovered thanks to large outbursts. They are found to be inactive or weakly active before the orbit of discovery, and are still active in the proceeding orbit of the reactivation. The reactivation is likely triggered by large nucleus disturbance or disruption that breaks the regolith that used to seal off the volatile, allowing sublimation-driven activity to resume. We found the rate of reactivation for short-period comets to be $0.007~\mathrm{comet^{-1}~orbit^{-1}}$, implying typical short-period comets only become temporary dormant for less than a few times.

The small sample size makes it difficult to interpret the findings. For example, it is unclear whether large outbursts are common in marking the reactivation of comets, and what mechanism causes such outburst. The recent research on 332P/Ikeya-Murakami signals that rotational instability may play an important role in reactivating small comets. It would be advisable to pay more attention on the comets that were discovered due to large outbursts, the most prominent ones being P/2010 H2 (Vales) and P/2013 YG46 (Spacewatch), as well as the unsolved case of 297P/Beshore.

The recognizability-rotational instability analysis also suggests that active sub-km-sized comets are quickly eliminated due to rotational instability before current NEO surveys can find them. Next generation high-cadence surveys, such as Asteroid Terrestrial-impact Last Alert System \citep[ATLAS;][]{Denneau2016}, Zwicky Transient Facility \citep[ZTF;][]{Ye2017} and Large Synoptic Survey Telescope \citep[LSST;][]{Collaboration2009}, are likely to find these short-lived comets before they are gone.

\acknowledgments

I thank an anonymous referee and Man-To Hui for helpful comments, David Clark for discussion about archival data search, Davide Farnocchia and Gareth Williams for discussion about the validity of the pre-discovery observations, as well as Eric Christensen and Robert Seaman for their help with the Catalina Sky Survey (CSS) and Siding Spring Survey (SSS) data. I also thank the support of the GROWTH project, funded by the National Science Foundation under Grant No. 1545949. The SkyMorph service was developed under NASA's Applied Information Systems Research (AISR) program. The Solar System Object Search service is hosted at the Canadian Astronomy Data Centre operated by the National Research Council of Canada with the support of the Canadian Space Agency. The pre-discovery image of 297P/Beshore was obtained from the Isaac Newton Group Archive which is maintained as part of the CASU Astronomical Data Centre at the Institute of Astronomy, Cambridge. The SSS survey was operated by the CSS in collaboration with the Australian National University. The CSS/SSS surveys are funded by the National Aeronautics and Space Administration under grant No. NNX15AF79G-NEOO, issued through the Science Mission Directorate's Near Earth Object Observations Program. This research has made use of data and/or services provided by the International Astronomical Union's Minor Planet Center.

\bibliographystyle{apj}
\bibliography{man}

\clearpage

\clearpage

\clearpage

\begin{table}
\begin{center}
\caption{Comets that were first detected in or after 2005 and have been observed for at least 2 orbits as of 2016 December 16. Comets with pre-discovery images identified in this work and \citet{Hui2016a} are highlighted.\label{tbl:all}}
\begin{tabular}{lll}
 \hline
 \textbf{213P/Van Ness} & 233P/La Sagra & 238P/Read \\
 249P/LINEAR & 255P/Levy & 257P/Catalina \\
 259P/Garradd & 260P/McNaught & 261P/Larson \\
 263P/Gibbs & \textbf{266P/Christensen} & 267P/LONEOS \\
 277P/LINEAR & 278P/McNaught & 284P/McNaught \\
 286P/Christensen & 287P/Christensen & \textbf{293P/Spacewatch} \\
 294P/LINEAR & \textbf{297P/Beshore} & 298P/Christensen \\
 300P/Catalina & \textbf{302P/Lemmon-PANSTARRS} & 309P/LINEAR \\
 310P/Hill & 316P/LONEOS-Christensen & \textbf{317P/WISE} \\
 319P/Catalina-McNaught & 325P/Yang-Gao & \textbf{332P/Ikeya-Murakami} \\
 333P/LINEAR & 335P/Gibbs & \textbf{336P/McNaught} \\
 337P/WISE & 338P/McNaught & 339P/Gibbs \\
 340P/Boattini & 341P/Gibbs & 345P/LINEAR \\
 P/2008 Y12 (SOHO) & & \\
 \hline
\end{tabular}
\end{center}
\end{table}

\clearpage

\begin{deluxetable}{lccccccl}
\rotate
\tablecaption{Pre-discovery (non-)detection of the comets of interests, including observational date and facility, heliocentric distance ($r_\mathrm{h}$), predicted visual total magnitude ($m_\mathrm{T, p}$) using the relation derived by JPL database, observed visual total magnitude (or upper limit; $m_\mathrm{T, o}$), and positional error. All magnitudes are in Johnson $V$.\label{tbl:can}}
\tablewidth{0pt}
\tablehead{
\colhead{Comet} & \colhead{Date} & \colhead{Facility} & \colhead{$r_\mathrm{h}$ (AU)} & \colhead{$m_\mathrm{T, p}$} & \colhead{$m_\mathrm{T, o}$} & \colhead{Pos. error\tablenotemark{a}} & \colhead{Note} }
\startdata
213P/Van Ness\tablenotemark{b} & 2002 Jan. 6 & NEAT & $4.72$ & $20.6$ & $>20.0$ & $1.4^\circ$ & \\
.. & 2002 Feb. 5 & NEAT & $4.71$ & $20.6$ & $>20.0$ & $1.5^\circ$ & \\
.. & 2002 Feb. 15 & NEAT & $4.71$ & $20.6$ & $>19.0$ & $1.5^\circ$ & Bright background \\
.. & 2003 Apr. 16 & NEAT & $4.11$ & $19.6$ & $>20.0$ & $0.9^\circ$ & \\
.. & 2003 Apr. 25 & NEAT & $4.09$ & $19.7$ & $>19.0$ & $0.9^\circ$ & Trailed image \\
.. & 2003 May 6 & NEAT & $4.06$ & $19.7$ & $>20.0$ & $0.8^\circ$ & \\
.. & 2003 May 14 & NEAT & $4.04$ & $19.7$ & $>20.0$ & $0.8^\circ$ & Star interference \\
266P/Christensen & 2001 Apr. 4 & NEAT & $2.98$ & $19.7$ & $>18.5$ & $79''$ & \\
.. & 2001 Apr. 26 & NEAT & $3.05$ & $20.0$ & $>18.5$ & $72''$ & \\
297P/Beshore & 2001 Mar. 20 & INT & $2.65$ & $\sim22$\tablenotemark{c} & $16.6$ & $11''$ & Detected; trailed \\
.. & 2001 Mar. 24 & NEAT & $2.64$ & $>18.5$ & $16.5$ & $11''$ & Bright background \\
.. & 2001 Apr. 22 & NEAT & $2.58$ & $>20.0$ & $16.4$ & $10''$ & \\
.. & 2002 May 25 & NEAT & $2.82$ & $>20.0$ & $17.9$ & $3''$ & \\
.. & 2002 May 26 & NEAT & $2.83$ & $>19.5$ & $17.9$ & $3''$ & \\
.. & 2002 Jun. 7 & NEAT & $2.85$ & $>18.0$ & $17.9$ & $3''$ & Trailed image \\
.. & 2002 Jul. 15 & NEAT & $2.96$ & $>20.0$ & $18.0$ & $4''$ & \\
.. & 2002 Jul. 16 & NEAT & $2.96$ & $>20.0$ & $18.0$ & $4''$ & \\
.. & 2002 Aug. 11 & NEAT & $3.04$ & $>20.0$ & $18.3$ & $4''$ & \\
.. & 2002 Aug. 30 & NEAT & $3.09$ & $>20.0$ & $18.6$ & $4''$ & \\
.. & 2007 May 16 & CSS & $2.99$ & $>19.5$ & $19.0$ & $5''$ & \\
.. & 2008 Mar. 5 & SSS & $2.41$ & $>19.5$ & $16.1$ & $1''$ & \\
302P/Lemmon-PANSTARRS & 1998 Aug. 18 & NEAT & $3.60$ & $19.2$ & $>18.5$ & $28''$ & \\
317P/WISE & 2005 Apr. 15 & CSS & $1.54$ & $20.0$ & $>19.0$ & $3''$ & \\
.. & 2005 May 8 & CSS & $1.40$ & $19.4$ & $>19.0$ & $3''$ & \\
.. & 2005 Jun. 1 & SSS & $1.27$ & $18.9$ & $>18.5$ & $2''$ & \\
.. & 2005 Jul. 28 & SSS & $1.22$ & $18.5$ & $\sim 19$ & $3''$ & Detected \\
.. & 2005 Aug. 16 & SSS & $1.29$ & $19.1$ & $>18.5$ & $3''$ & \\
332P/Ikeya-Murakami\tablenotemark{d} & 2003 Sep. 25 & CFHT & $4.05$ & $21.0$ & $>22.9$\tablenotemark{e} & $0.8^\circ$ & \\
.. & 2003 Sep. 27 & CFHT & $4.05$ & $21.0$ & $>23.4$\tablenotemark{e} & $0.8^\circ$ & Partial coverage \\
.. & 2005 Apr. 19 & CSS & $1.60$ & $10.6$ & $>19.5$ & $2^\circ$ & \\
.. & 2005 Apr. 30 & CSS & $1.59$ & $10.6$ & $>19.5$ & $2^\circ$ & \\
336P/McNaught & 1996 Aug. 9 & NEAT & $2.85$ & $18.4$ & $>19.0$ & $6''$ & Star interference \\
.. & 1996 Aug. 11 & NEAT & $2.85$ & $18.4$ & $19.5$ & $6''$ & Detected \\
.. & 2005 May 10 & WHT & $3.98$ & $22.6$ & $>20.5$ & $5''$ & \\
.. & 2006 Feb. 8 & SSS & $2.95$ & $19.4$ & $>19.0$ & $1''$ & \\
.. & 2006 Mar. 23 & SSS & $2.83$ & $18.5$ & $>19.0$ & $1''$ & Star interference \\
\enddata
\tablecomments{Abbreviation of surveys/facilities: CFHT - Canada-France-Hawaii Telescope; CSS - Catalina Sky Survey; INT - Issac Newton Telescope; NEAT - Near-Earth Asteroid Tracking; SSS - Siding Spring Survey; WHT = William Herschel Telescope}
\tablenotetext{a}{Angular width of the $3\sigma$ error ellipse semimajor axis provided by the JPL database.}
\tablenotetext{b}{The predicted magnitude may be errornous; orbital uncertainty is calculated by the author instead of retrieving from the JPL database. See main text.}
\tablenotetext{c}{The comet is trailed; the reported brightness has been corrected for trailing loss.}
\tablenotetext{d}{Data recalculated from \citet{Hui2016a}.}
\tablenotetext{e}{Magnitudes are converted to Johnson $V$ using the transformation equation derived by \citet{2005AJ....130..873J}.}
\end{deluxetable}

\clearpage

\begin{deluxetable}{lllll}
\tabletypesize{\scriptsize}
\rotate
\tablecaption{Facilities involved in the archival observations in Table~\ref{tbl:can}.\label{tbl:fac}}
\tablewidth{0pt}
\tablehead{
\colhead{Facility} & \colhead{Location} & \colhead{Telescope} & \colhead{Field of view} & \colhead{Image resolution} }
\startdata
 CFHT & Maunakea, Hawai'i, U.S.A. & 3.6-m reflecting telescope + MegaCam & 1 deg$^2$ & 0.2''/pixel \\
 CSS & Mt. Catalina, Arizona, U.S.A. & 0.68-m Schmidt telescope & 8 deg$^2$ & 2.5''/pixel \\
 INT & La Palma, Canary Islands, Spain & 2.5-m Isaac Newton Telescope + WFC & 0.3 deg$^2$ & 0.3''/pixel \\
 NEAT (1995--2000) & Haleakala, Hawai'i, U.S.A. & 1.0-m GEODSS telescope & 2 deg$^2$ & 1''/pixel \\
 NEAT (2000--2003) & Maui, Hawai'i, U.S.A. & 1.2-m AMOS telescope & 2 deg$^2$ & 1''/pixel \\
 NEAT (2001--2007) & Palomar Mountain, California, U.S.A. & 1.2-m Oschin Schmidt & 5 deg$^2$ & 1''/pixel \\
 SSS & Siding Spring Observatory, Australia & 0.5-m Uppsala Schmidt & 4 deg$^2$ & 2''/pixel \\
 WHT & La Palma, Canary Islands, Spain & 4.2-m William Herschel Telescope + Prime focus imager & 0.07 deg$^2$ & 0.2''/pixel \\
\enddata
\end{deluxetable}

\clearpage

\begin{table}
	\begin{center}
		\caption{Original orbit of 213P/Van Ness from JPL orbit \#67 versus the orbit derived from pre-outburst data only.\label{tbl:213p-orb}}
		\begin{tabular}{lll}
			\tableline\tableline
			& JPL \#67 & Pre-outburst orbit \\
			\tableline
			Epoch & 2012 Jun. 2.0 (TT) & 2012 Jun. 2.0 (TT) \\
			Perihelion time $T$ & 2011 Jun. 16.60716 (TT) & 2011 Jun. 9.25791 (TT) \\
			Perihelion distance $q$ (AU) & 3.4268640 & 3.4196637 \\
			Eccentricity $e$ & 0.3806647 & 0.3800714 \\
			Inclination $i$ (J2000.0) & $10.23692^\circ$ & $10.23292^\circ$ \\
			Longitude of the ascending node $\Omega$ (J2000.0) & $312.56369^\circ$ & $312.50968^\circ$ \\
			Argument of perihelion $\omega$ (J2000.0) & $3.51416^\circ$ & $3.41276^\circ$ \\
			Mean anomaly $M$ & $54.59485^\circ$ & $55.91280^\circ$ \\
			First observation & 2005 August 4 & 2005 August 4 \\
			Last observation & 2012 February 3 & 2005 August 31 \\
			Observations used & 3090 & 14 \\
			\tableline
		\end{tabular}
	\end{center}
\end{table}

\clearpage

\begin{table}
	\begin{center}
		\caption{Original orbit of 297P/Beshore from JPL orbit \#33 versus the updated orbit with pre-discovery observations.\label{tbl:297p-orb}}
		\begin{tabular}{lll}
			\tableline\tableline
			& JPL \#33 & Updated orbit \\
			\tableline
			Epoch & 2009 Mar. 8.0 (TT) & 2009 Mar. 8.0 (TT) \\
			Perihelion time $T$ & 2008 Mar. 21.01997 (TT) & 2008 Mar. 21.01638 (TT) \\
			Perihelion distance $q$ (AU) & 2.4086462 & 2.4086958 \\
			Eccentricity $e$ & 0.3086498 & 0.3086364 \\
			Inclination $i$ (J2000.0) & $10.26285^\circ$ & $10.26298^\circ$ \\
			Longitude of the ascending node $\Omega$ (J2000.0) & $98.28318^\circ$ & $98.28353^\circ$ \\
			Argument of perihelion $\omega$ (J2000.0) & $131.81419^\circ$ & $131.81361^\circ$ \\
			Mean anomaly $M$ & $53.34699^\circ$ & $53.34740^\circ$ \\
			First observation & 2008 May 6 & 2001 March 20 \\
			Last observation & 2014 June 18 & 2014 June 18 \\
			Observations used & 568 & 583 \\
			\tableline
		\end{tabular}
	\end{center}
\end{table}

\end{CJK*}
\end{document}